\documentclass[conference]{IEEEtran}
\usepackage{amsmath,amssymb,amsthm}
\usepackage{graphicx}
\usepackage{graphics}
\usepackage{epstopdf}
\usepackage{mathtools}
\usepackage{subfigure}
\usepackage{algorithm,algorithmic}

\newtheorem{remark}{Remark}

\newtheorem{theo}{Theorem}

\ifCLASSINFOpdf
\else
\fi

\begin{document}

\title{Pilot Pattern Adaptation for 5G MU-MIMO Wireless Communications}

\author{Nassar Ksairi, Beatrice Tomasi, and Stefano Tomasin\\
Mathematical and Algorithmic Sciences Lab, France Research Center,\\
Huawei Technologies Co. Ltd., Boulogne-Billancourt, France.\\
Emails:
\{nassar.ksairi, beatrice.tomasi, stefano.tomasin\}@huawei.com,
}

\IEEEoverridecommandlockouts
\IEEEpubid{\makebox[\columnwidth]{\hfill 978-1-5090-1749-2/16/\$31.00~\copyright~2016~IEEE} \hspace{\columnsep}\makebox[\columnwidth]{ }}

\maketitle

\begin{abstract}
To meet the goal of ten-fold increase in spectral efficiency, multiuser
multiple-input-multiple-output (MU-MIMO) techniques capable of achieving
high spatial multiplexing gains are expected to be an essential
component of fifth-generation (5G) radio access systems. This increase in
multiplexing gain, made possible by equipping base stations with a large number
of antennas, entails a proportional increase in channel state information (CSI)
acquisition overhead. This article addresses the problem of reducing this CSI
overhead by optimizing the amount of time-frequency resources allocated for
channel training purposes while not affecting the quality of the associated
channel estimate. First we show that in MU-MIMO, adapting pilot symbol density
in the time-frequency grid should be performed both on a per
resource block (RB) basis and on the basis of groups of users sharing similar
channel conditions. Next, we propose a practical scheme that can perform
grouping based per-RB pilot pattern adaptation.
Finally, we evaluate using both analytical and numerical results the gain in
spectral efficiency that can be achieved using this scheme as compared to
conventional MU-MIMO systems that use fixed pilot patterns.

\end{abstract}

\IEEEpeerreviewmaketitle

\section{Introduction}
\label{sec:intro}
In contrast to existing wireless systems, next-generation MU-MIMO will most
probably be deployed using base stations (BS) that are equipped with a large
number of antennas thus increasing the system spectral efficiency.
However, achieving this increase in spectral efficiency is conditioned on the
availability of precise estimates of the channels between the different users
and the BS~\cite{mimo_myths}. 
CSI is typically obtained by sending reference signals (RS), also called pilots,
which are known at both transmitter and receiver sides.
The portion of time and frequency resources reserved to these training sequences
is what constitutes the channel training overhead. In an uplink scenario
where multiple users are simultaneously transmitting to
the BS, the channel training overhead typically grows with the number of
these users. This also applies to downlink user-specific pilots transmitted
by the BS\footnote{As opposed to cell-specific reference signals, user-specific
pilots are transmitted only on the RBs on which the intended user is scheduled
and they pass through the same MIMO precoding applied to the data symbols.}.
There is thus a crucial need to compensate this increase in overhead. Whether in
the uplink or in the downlink, special care should be paid so that the pilot
signals of users scheduled at the same time-frequency resources are (at least
partially) orthogonal to each other and that the symbol positions used by one of
them for pilot transmission are not used by another for data transmission to
avoid data-pilot interference.

The issue of reducing channel training overhead in MU-MIMO was addressed in
\cite{beatrice_maxime} and \cite{jsdm} where the authors propose to exploit the
spatial correlation of users' channels to the BS antenna array in order to
minimize the length of their training sequences. The proposed schemes rely on
the fact that users' pilot signals can be separated to some extent at the BS
through their spatial signature, provided that their low-rank channel covariance
matrices are known.

Another promising approach to reduce channel training overhead without 
the additional overhead needed for spatial covariance estimation
consists in reducing the {\it average} density of pilot symbols. Indeed, in
wireless systems that use orthogonal frequency division multiplexing (OFDM),
channel training is done by sending pilots on some predefined positions, i.e.
according to a predefined {\it pilot pattern}, in the time-frequency resource
grid. Once an estimate of the channel at the pilot positions is available,
interpolation techniques are used to exploit correlation in time and
frequency, and obtain the estimate of the channel on the grid
positions carrying data. In principle, the required density of pilot symbols in
a pilot pattern is related to the level of correlation of the channel
coefficients along the time and the frequency axes. For instance, for users with
a fast changing channel, density along the time axis must be increased with
respect to (w.r.t.) almost-static users. On the other hand, a higher frequency
selectivity requires denser pilots along the frequency axis. Adapting the pilot
patterns to users' second-order statistics makes it possible to send/receive
training sequences with different pilot symbol densities: some of these can be
lower than the highest pilot density designed to cope with the worst-case
channel. In \cite{su_ofdm1}, methods for choosing the pilot pattern
for OFDM based on the channel time and frequency correlation
properties are proposed. A method for selecting MIMO OFDM pilot patterns based
on the channel signal-to-noise ratio (SNR), maximum Doppler frequency and root
mean square delay spread is proposed in~\cite{su_mimo}. In \cite{qos_pilot},
adaptive pilot patterns are proposed but the adaptation is done only w.r.t.
users' quality-of-service (QoS) requirements.
A method to assign OFDMA pilot patterns on the basis of groups of mobile
users having the same speed is proposed in \cite{speed_pilot}. However, in this
method groups with different pilot patterns are forced to occupy disjoint time
intervals. All these works do not address the more challenging issue of pilot
pattern adaptation for MU-MIMO systems where the pilot symbols of different
users could overlap due to spatial multiplexing.

\subsection*{Contributions}
 We propose a pilot pattern adaptation scheme that can lower pilot and
signaling overhead for both uplink and downlink user-specific training sequences
in MU-MIMO systems. The scheme consists in constraining the scheduler to group
users based on their channels second-order statistics. Therefore, patterns with
a reduced overhead can be used on a RB in which all scheduled users have milder
requirements on pilots. Even though the scheme constrains the scheduler
with the grouping step, we prove that the average spectral efficiency achieved
with the proposed scheme is guaranteed to be larger than that of conventional
pilot selection combined with any user scheduling paradigm, provided that the
number of BS antennas and of cell users is large enough.
Finally, we show through simulations that this property is valid even with
practical values of the number of BS antennas and of cell users.

\section{System Model}
\label{sec:model}

We consider an OFDM-based MU-MIMO single-cell system where the BS is equipped
with $M\gg1$ antennas and assume that the OFDM resource grid is divided into
$N_{\mathrm{RB}}>1$ RBs, each composed of $N_s$ OFDM symbols, each comprising
$N_{\mathrm{SC}}$ subcarriers (SCs), resulting in a total of
$N_{\mathrm{RE}}=N_s\times N_{\mathrm{SC}}$ resource elements (REs) per RB.
We denote the set of (single-antenna) terminals asking to be served as
$\mathcal{K}$ and define $K\stackrel{\mathrm{def}}{=}|\mathcal{K}|$.
Let us focus on RB~$r$ ($r\in\{1\cdots N_{\mathrm{RB}}\}$) and let
$\mathcal{U}_r^{\mathrm{UL}}\subset\mathcal{K}$
(resp. $\mathcal{U}_r^{\mathrm{DL}}$) designate the set of users assigned to
this RB for uplink (resp. downlink) transmission such that
\begin{equation}
 \label{eq:mux}
 \max_{r\in\{1\cdots N_{\mathrm{RB}}\}}
\left\{\left|\mathcal{U}_r^{\mathrm{UL}}\right|,
\left|\mathcal{U}_r^{\mathrm{DL}}\right|\right\}\leq U^{\mathrm{mux}},
\end{equation}
where $U^{\mathrm{mux}}$ is the maximum spatial multiplexing gain allowed by the
system.
Define $\mathcal{D}^{\mathrm{UL}}$ and $\mathcal{D}^{\mathrm{DL}}$ as the
subsets of $\{1\cdots N_s\}\times\{1\cdots N_{\mathrm{SC}}\}$ that are
used for uplink and downlink data transmission, respectively.
Similarly, define $\mathcal{P}^{\mathrm{UL}}$ and $\mathcal{P}^{\mathrm{DL}}$
as the associated subsets of REs used for pilot transmission.
The division of the set of REs in one RB into $\mathcal{D}^{\mathrm{UL}}$ and
$\mathcal{P}^{\mathrm{UL}}$ (or into $\mathcal{D}^{\mathrm{DL}}$ and
$\mathcal{P}^{\mathrm{DL}}$) is typically dictated by the so-called
{\it pilot pattern} defined by the communications standard.
Finally, denote by $\mathbf{h}_{k,r,t,n}^{\mathrm{UL}}$ and
$\mathbf{h}_{k,r,t,n}^{\mathrm{DL}}$ the vector of
small-scale fading coefficients at subcarrier $n$ 
($n\in\{1\cdots N_{\mathrm{SC}}\}$) during the $t$-th OFDM symbol
($t\in\{1\cdots N_s\}$) from user~$k\in\mathcal{K}$ to the $M$ antenna elements
at the BS and from these antennas to user~$k$, respectively.
The samples $\mathbf{y}_{r,t,n}$ and $y_{k,r,t,n}$ received
respectively at the BS and by user~$k$ are given by
\begin{align}
 \mathbf{y}_{r,t,n}&=\sum_{k\in\mathcal{U}_r^{\mathrm{UL}}}
\sqrt{\eta_k P^{\mathrm{UL}}}\mathbf{h}_{k,r,t,n}^{\mathrm{UL}}x_{k,r,t,n}+
\mathbf{v}_{r,t,n}\:,\label{eq:ul_samples}\\
 y_{k,r,t,n}&=\sqrt{\eta_k P^{\mathrm{DL}}}
\left(\mathbf{h}_{k,r,t,n}^{\mathrm{DL}}\right)^{\mathrm{T}}
\mathbf{x}_{r,t,n}+v_{k,r,t,n}\:,\label{eq:dl_samples}
\end{align}
where $\mathbf{v}_{r,t,n}$ and $v_{k,r,t,n}$
are independent identically-distributed (i.i.d.)
$\mathcal{CN}\left(0,\sigma^2\right)$ noise samples, $\eta_k$ is the large-scale
fading factor, $P^{\mathrm{UL}}$ is the users' transmit power and
$P^{\mathrm{DL}}$ is the transmit power of the BS. As for $x_{k,r,t,n}$ and
$\mathbf{x}_{r,t,n}$, they are zero-mean unit-power symbols sent by user~$k$ and
the BS, respectively. In the sequel, we use the notations
$\{s_{k,r,t,n}\}_{(t,n)\in\mathcal{D}}$ and
$\{p_{k,r,t,n}\}_{(t,n)\in\mathcal{P}}$ to designate respectively the set of
data symbols and of pilot symbols in RB~$r$:
\begin{equation}
 \begin{multlined}
 \forall k\in\mathcal{U}_r^{\mathrm{UL}}, x_{k,r,t,n}=
\left\{
\begin{array}{ll}
 s_{k,r,t,n},& (t,n)\in \mathcal{D}^{\mathrm{UL}},\\
 p_{k,r,t,n},& (t,n)\in \mathcal{P}^{\mathrm{UL}}.
\end{array}
\right.
 \end{multlined}
\end{equation}
In the uplink, we assume that {\it linear combining} is used to detect
users' signals based on the samples
$\frac{1}{M}\left(\mathbf{w}_{k,r,t,n}^{\mathrm{UL}}\right)^{\mathrm{H}}
\mathbf{y}_{r,t,n}$, where $\mathbf{w}_{k,r,t,n}^{\mathrm{UL}}$ is the 
combining vector for user $k\in\mathcal{U}_r^{\mathrm{UL}}$. These combining
vectors are typically chosen depending on $\mathcal{U}_r^{\mathrm{UL}}$ through
some optimality criteria such as maximum-ratio combining (MRC) and zero-forcing
(ZF) combining. Similarly, we assume that the BS applies {\it linear precoding}
in the downlink so that
\begin{equation}
  \mathbf{x}_{r,t,n}=
\left\{
\begin{array}{ll}\sum_{k\in\mathcal{U}_r^{\mathrm{DL}}}\frac{1}{M}
\mathbf{w}_{k,r,t,n}^{\mathrm{DL}}s_{k,r,t,n},& (t,n)\in
\mathcal{D}^{\mathrm{DL}},\\
\sum_{k\in\mathcal{U}_r^{\mathrm{DL}}}\frac{1}{M}
\mathbf{w}_{k,r,t,n}^{\mathrm{DL}}p_{k,r,t,n},& (t,n)\in
\mathcal{P}^{\mathrm{DL}},
\end{array}\right.
\end{equation}
where $\mathbf{w}_{k,r,t,n}^{\mathrm{DL}}$ is the precoding vector assigned to
user $k\in\mathcal{U}_r^{\mathrm{DL}}$ and normalized in accordance with
$P^{\mathrm{DL}}$. Here, $p_{k,r,t,n}$ is a
{\it user-specific} pilot symbol that undergoes the same precoding as
the data symbol $s_{k,r,t,n}$ and which is intended for the estimation
of the effective channel $\frac{1}{M}
\left(\mathbf{h}_{k,r,t,n}^{\mathrm{DL}}\right)^{\mathrm{T}}
\mathbf{w}_{k,r,t,n}^{\mathrm{DL}}$ at the user terminal.
Vectors $\mathbf{w}_{k,r,t,n}^{\mathrm{DL}}$ are typically based on
$\mathcal{U}_r^{\mathrm{DL}}$ using some optimality criteria, e.g.
maximum-ratio transmission (MRT) and zero-forcing (ZF) precoding.

Each entry of $\mathbf{h}_{k,r,t,n}^{\mathrm{UL}}$ and
$\mathbf{h}_{k,r,t,n}^{\mathrm{DL}}$ is assumed to be a two-dimensional
wide-sense stationary (WSS) random process that is band
limited~\cite{pilot_rule} w.r.t. both $t$ and $n$.
In other words, the Fourier transform of both its $t$-axis and its $n$-axis
auto-correlation functions has a finite support. The highest value in the
frequency domain support is the maximum Doppler frequency
shift denoted as $f_k^D$, while the largest value in the time domain
support is the maximum delay spread denoted as $\tau_k^{\max}$.
We assume that $\forall k\in\mathcal{K}$,
the pair $(\tau_k^{\max},f_k^D)$ can take only a finite number $G>1$ of values
denoted as $\{(\tau_g,f_g)\}_{1\leq g\leq G}$. In practice, this assumption
amounts to quantizing the different values of $(\tau_k^{\max},f_k^D)$.
The set of users whose channels follows the $g$-th model are denoted as
$\mathcal{G}_g$, where
\begin{equation}
 \label{eq:group_def}
\mathcal{G}_g\stackrel{\mathrm{def}}{=}\left\{k\in\mathcal{K}|
(\tau_k^{\max},f_k^D)=(\tau_g,f_g)\right\}, 1\leq g\leq G.
\end{equation}
As in~\cite{pilot_rule}, we assume that one can get
small-enough
\footnote{In the sense that the associated channel
estimation MSE does not exceed a predefined target value.}
channel estimate mean-square error (MSE) by restricting
$\mathcal{P}^{\mathrm{UL}}$ and $\mathcal{P}^{\mathrm{DL}}$ to be composed of
regularly spaced positions with a pilot symbol density two-times the density
dictated by the {\it sampling theorem for band limited WSS random processes}.
This rule of thumb implies that the {\it maximum} pilot symbol spacing that can
be used on a channel~$g$ is $\Delta_g^s$ OFDM symbols in the time domain and
$\Delta_g^{\mathrm{SC}}$ SCs in the frequency domain, where
\begin{equation}
 \label{eq:delta_npg}
 \Delta_g^s\stackrel{\mathrm{def}}{=}
\left\lfloor\frac{1}{4f_g T_s}\right\rfloor,
\quad\Delta_g^{\mathrm{SC}}\stackrel{\mathrm{def}}{=}
\left\lfloor \frac{1}{4\tau_g \Delta f}\right\rfloor.
\end{equation}
Here, $T_s$ denotes the duration of the OFDM symbol and $\Delta f$ the
inter-subcarrier frequency separation.

\section{Conventional MU-MIMO Pilot Patterns}
\label{sec:conventional}

\begin{figure}[h]
 \centering
 \subfigure{
	\centering
  \includegraphics[width=0.45\linewidth]{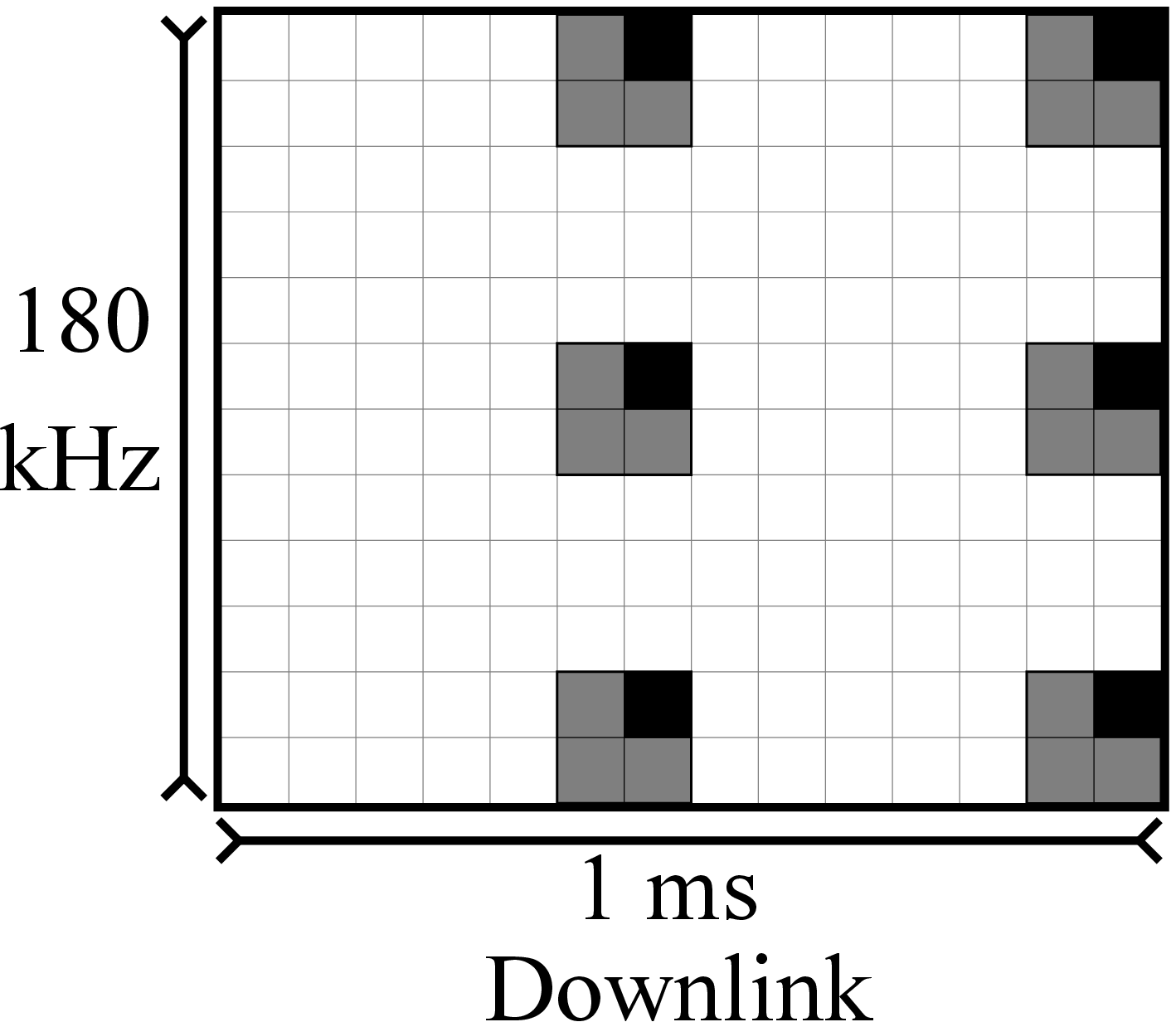}
 }
 ~
 \subfigure{
	\centering
  \includegraphics[width=0.45\linewidth]{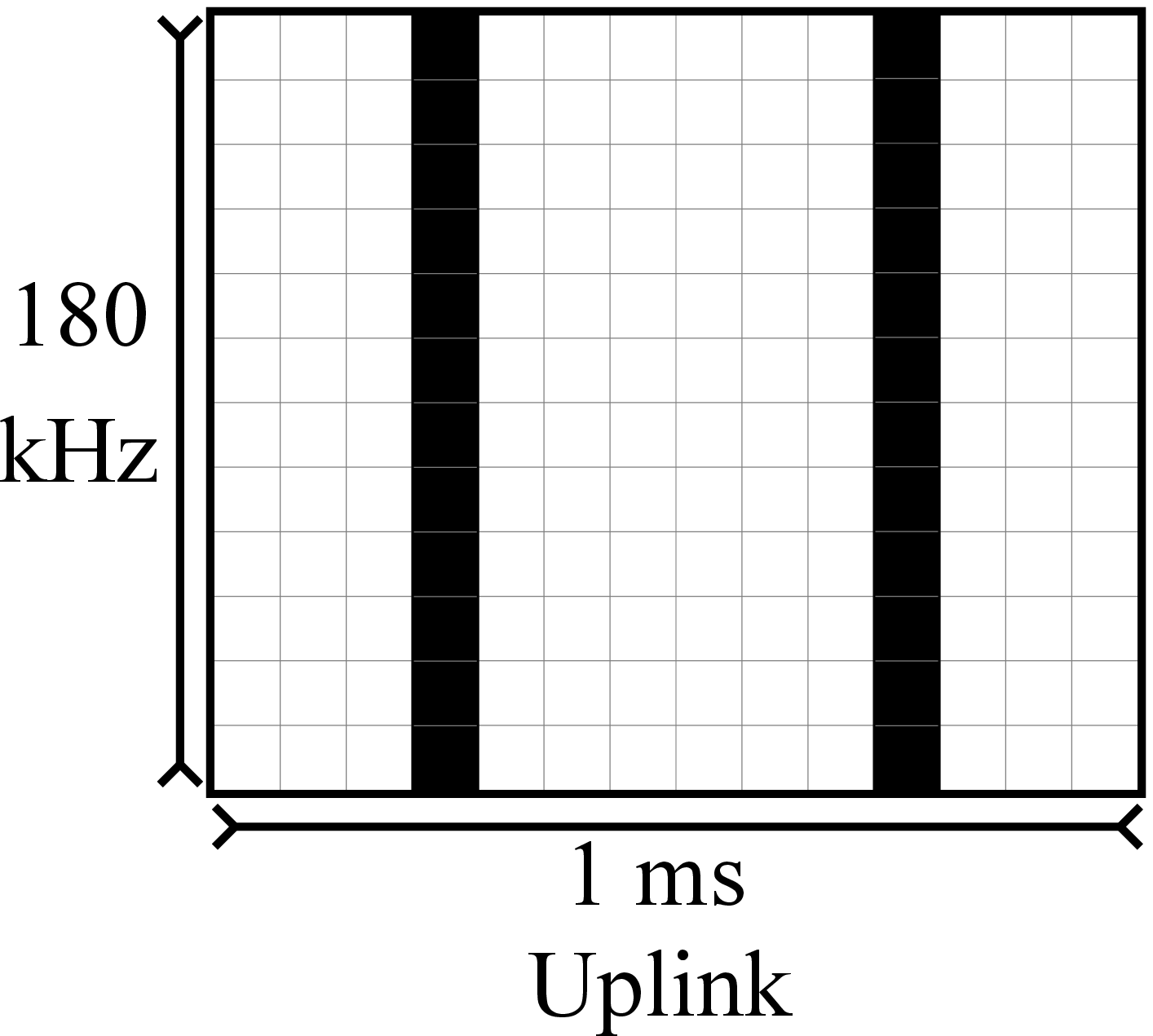}
 }
 \caption{Pilot pattern for 4-layer user-specific RS in LTE-Advanced}
 \label{fig:dl_ul_pattern_with_mux}
\end{figure}

In current wireless systems, the same pilot pattern, denoted as
$\mathcal{P}^{\mathrm{conv},\mathrm{UL}}$
($\mathrm{conv}$ stands for `conventional'), is
used on all uplink RBs while the same pilot pattern, denoted as
$\mathcal{P}^{\mathrm{conv},\mathrm{DL}}$, is
used on all downlink RBs. Both $\mathcal{P}^{\mathrm{conv},\mathrm{UL}}$
and $\mathcal{P}^{\mathrm{conv},\mathrm{DL}}$ are designed to cope with
the worst-case scenario in which $\forall r,g$,
$\mathcal{U}_r^{\mathrm{UL}}\cap\mathcal{G}_g\neq\emptyset$ and
$\mathcal{U}_r^{\mathrm{DL}}\cap\mathcal{G}_g\neq\emptyset$. Combining this with
the requirement that the total number of pilot symbols in a RB should be an
integer multiple of the number of multiplexed users, we get
\begin{equation}
 \label{eq:worst_case_p}
 \left|\mathcal{P}^{\mathrm{conv},\mathrm{UL}}\right|=
\left|\mathcal{P}^{\mathrm{conv},\mathrm{DL}}\right|=
\max_{g\in\{1\cdots G\}}
\left\lceil\frac{N_s}{\Delta_g^s}\right\rceil
\left\lceil\frac{N_{\mathrm{SC}}}{\Delta_g^{\mathrm{SC}}}\right\rceil
U^{\mathrm{mux}}
\end{equation}
Plugging $U^{\mathrm{mux}}=4$ and the LTE system parameters
into~\eqref{eq:worst_case_p} while assuming a worst-case Doppler frequency shift
$f^D = 300$ Hz and maximum delay spread $\tau^{\max}=4.69$ ms
yields $\left|\mathcal{P}^{\mathrm{conv}}\right|=24$, in agreement with the
uplink and downlink pilot patterns of LTE-Advanced shown in
Fig.~\ref{fig:dl_ul_pattern_with_mux}.

The average spectral efficiency $R_r$ of RB $r$ on which the subset
$\mathcal{U}_r$ of users is scheduled with pilot pattern $\mathcal{P}_r$ is
\begin{equation}
 \label{eq:general_R}
 R_{r}\left(\mathcal{U}_r,\mathcal{P}_r\right)
\stackrel{\mathrm{def}}{=}
\frac{1}{N_{\mathrm{RE}}}
\sum_{k\in\mathcal{U}_r}\sum_{(t,n)\notin\mathcal{P}_r}
\log\left(1+\mathrm{SINR}_{k,r,t,n}\right)
\end{equation}
where $\log$ is the base-2 logarithm and where $\mathrm{SINR}_{k,r,t,n}$
is user $k$ signal-to-interference-plus-noise ratio on $(r,t,n)$ that is given
by $\mathrm{SINR}_{k,r,t,n}=\mathrm{SINR}_{k,r,t,n}^{\mathrm{UL}}$ in the uplink
and by $\mathrm{SINR}_{k,r,t,n}=\mathrm{SINR}_{k,r,t,n}^{\mathrm{DL}}$ in the
downlink. Here, we defined
\begin{equation}
 \label{eq:uplink_sinr}
 \begin{multlined}
  \mathrm{SINR}_{k,r,t,n}^{\mathrm{UL}}
\stackrel{\mathrm{def}}{=}\\
\frac{\eta_k P^{\mathrm{UL}}
|\mathbf{w}_{k,r,t,n}^{\mathrm{H}}\mathbf{h}_{k,r,t,n}|^2}{\sum_{j\neq k}
\eta_j P^{\mathrm{UL}}|\mathbf{w}_{k,r,t,n}^{\mathrm{H}}\mathbf{h}_{j,r,t,n}|^2+
\mathbf{w}_{k,r,t,n}^{\mathrm{H}}\mathbf{w}_{k,r,t,n}\sigma^2}\:.
 \end{multlined}
\end{equation}
and
\begin{equation}
 \label{eq:downlink_sinr}
 \begin{multlined}
  \mathrm{SINR}_{k,r,t,n}^{\mathrm{DL}}
\stackrel{\mathrm{def}}{=}\\
\frac{\eta_k P^{\mathrm{DL}}
|\mathbf{w}_{k,r,t,n}^{\mathrm{H}}\mathbf{h}_{k,r,t,n}|^2}{\sum_{j\neq k}
\eta_k P^{\mathrm{DL}}|\mathbf{w}_{j,r,t,n}^{\mathrm{H}}\mathbf{h}_{k,r,t,n}|^2+
M^2 \sigma^2}\:.
 \end{multlined}
\end{equation}
The associated maximum spectral efficiency is given by
\begin{equation}
 \label{eq:max_conv}
 R^{\mathrm{conv}}\stackrel{\mathrm{def}}{=}
\max_{\left\{\mathcal{U}_r\right\}_{r\in\{1\cdots	N_{\mathrm{RB}}\}}}
\frac{1}{N_{\mathrm{RB}}}\sum_{r=1}^{N_{\mathrm{RB}}}
R_{r}\left(\mathcal{U}_r,\mathcal{P}^{\mathrm{conv}}\right).
\end{equation}
Solving \eqref{eq:max_conv} involves high CSI acquisition
overhead needed to have CSI at the BS about each user's channel on all the RBs.
Many of the existing user scheduling paradigms try to find the RB
allocation that solves (exactly or approximately) \eqref{eq:max_conv}.
Finding such RB allocation is out of the scope of this work. However, we
evaluate how the proposed pilot allocation and grouping affect
the maximum spectral efficiency of the system.

For the sake of notational simplicity, we drop from now on the use of
superscripts $\mathrm{DL}$ and $\mathrm{UL}$.
For instance, the notations $\mathcal{U}_r^{\mathrm{UL}}$ and
$\mathcal{U}_r^{\mathrm{DL}}$ are merged into
$\mathcal{U}_r$ while $\mathcal{P}^{\mathrm{conv},\mathrm{UL}}$ and
$\mathcal{P}^{\mathrm{conv},\mathrm{UL}}$ are replaced with
$\mathcal{P}^{\mathrm{conv}}$.
Whenever needed, the transmission scenario, whether downlink or uplink, will be
explicitly mentioned.

\section{Adaptive Pilot Pattern Selection and User Grouping for MU-MIMO}
\label{sec:description}

Using the pilot patterns of Section~\ref{sec:conventional} for user-specific RS
in 5G systems would be very inefficient. Indeed, in
these systems: $i$) users' channel conditions in one cell can be very diverse
due to their larger numbers, and $ii$) longer pilot sequences are needed because
more users are spatially multiplexed. For example, a MU-MIMO transmission to 8
users would require 48 pilot symbols per RB as opposed to 24 in the case of 4
multiplexed users. We thus propose an {\it adaptive} pilot pattern selection
that is based on the following guidelines.

\subsection{Guidelines}
\label{sec:guidelines}

\begin{figure}[h]
 \centering
  \includegraphics[width=0.80\linewidth]{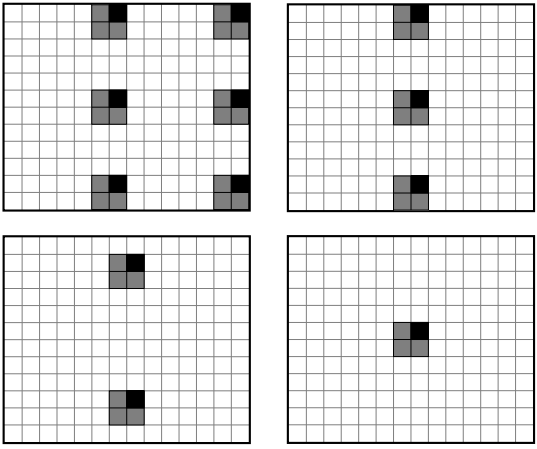}
 \caption{Example of adaptive pilot patterns for 4-layer user-specific RS.}
 \label{fig:dl_pattern_example}
\end{figure}

Let $\mathcal{R}\subset2^{\{1\cdots N_s\}\times\{1\cdots N_{\mathrm{SC}}\}}$
designate the predefined set of possible values of $\mathcal{P}$ and assume that
any $\mathcal{P}\in\mathcal{R}$ has a regular pilot symbol spacing denoted as
$(\delta_{\mathcal{P}}^s,\delta_{\mathcal{P}}^{\mathrm{SC}})$ which satisfies
\begin{equation}
 \label{eq:distinct_spacing}
 \forall\mathcal{P},\mathcal{Q}\in\mathcal{R}\textrm{ s.t. }
\mathcal{P}\neq\mathcal{Q},
(\delta_{\mathcal{P}}^s,\delta_{\mathcal{P}}^{\mathrm{SC}})\neq
(\delta_{\mathcal{Q}}^s,\delta_{\mathcal{Q}}^{\mathrm{SC}}).
\end{equation}
The constraint in \eqref{eq:distinct_spacing} means that for two pilot patterns
to be considered as distinct they should have different pilot symbol spacing
values, either on the time axis or on the frequency axis or on both.
It is also natural to bound the number of possible pilot patterns with the
number $G$ of distinct statistical channel conditions:
\begin{equation}
 \label{eq:npg_cond}
 N_{\mathcal{R}}\stackrel{\mathrm{def}}{=}\left|\mathcal{R}\right|\leq G\:.
\end{equation}
This paper focuses on the practical case where channel spatial covariance
matrices are not known at the BS and where, consequently, the training sequence
shortening techniques of \cite{beatrice_maxime} or~\cite{jsdm} do not apply.
We thus impose that,
\begin{equation}
 \label{eq:multiple_mux}
 \forall\mathcal{P}\in\mathcal{R},|\mathcal{P}|=
\left\lceil\frac{N_s}{\delta_{\mathcal{P}}^s}\right\rceil
\left\lceil\frac{N_{\mathrm{SC}}}{\delta_{\mathcal{P}}^{\mathrm{SC}}}
\right\rceil U^{\mathrm{mux}}.
\end{equation}
Figure~\ref{fig:dl_pattern_example} shows $N_{\mathcal{R}}=4$
pilot patterns which satisfy conditions
\eqref{eq:distinct_spacing}--\eqref{eq:multiple_mux}
for a system with $G\geq4$ and $U^{\mathrm{mux}}=4$.
In the following, we assume that such patterns can be used for both
uplink and downlink pilot transmissions.

We use the term {\it pilot pattern adaptation} to designate any
mapping $\mathcal{P}_r$ from $\{1\cdots N_{\mathrm{RB}}\}\times2^{\mathcal{K}}$
to $\mathcal{R}$ where $\mathcal{R}$ satisfies conditions
\eqref{eq:distinct_spacing}, \eqref{eq:npg_cond} and \eqref{eq:multiple_mux}.
Any such mapping assigns to each RB~$r$ used by a set $\mathcal{U}_r$ of users a
set of pilot positions $\mathcal{P}_r(\mathcal{U}_r)$ indicated as
$\mathcal{P}_r$. In other words, pilot pattern adaptation is performed on a
per-RB basis so that all users scheduled in the same RB have the same pilot
pattern, thus avoiding interference between data and pilot symbols.
Moreover, the pilot pattern on any RB should accommodate the user
with the worst-case statistical channel conditions scheduled in that RB,
i.e. $\forall r\in\{1\cdots N_{\mathrm{RB}}\}$,
\begin{align}
 \label{eq:guide2}
\delta_{\mathcal{P}_r}^{\mathrm{SC}}&\leq\min\left\{
\Delta_g^{\mathrm{SC}}|g\in\{1\cdots G\},
\mathcal{G}_g\cap\mathcal{U}_r\neq\emptyset\right\},\nonumber\\
\delta_{\mathcal{P}_r}^s&\leq\min\left\{\Delta_g^s|g\in\{1\cdots G\},
\mathcal{G}_g\cap\mathcal{U}_r\neq\emptyset\right\}.
\end{align}
Finally, scheduling should take into account users' pilot
density requirements, which implies the need for an additional process that 
identifies the groups $\mathcal{G}_1$, \dots, $\mathcal{G}_G$
and which interacts with the scheduling process, i.e.
\begin{equation}
 \label{eq:guide3}
  \forall r\in\{1\cdots N_{\mathrm{RB}}\},
\mathcal{U}_r=\mathcal{U}_r\left(\mathcal{G}_1,\ldots,\mathcal{G}_G\right).
\end{equation}

We propose a method that performs both pilot pattern
adaptation and user scheduling following the guidelines in
\eqref{eq:distinct_spacing}--\eqref{eq:guide3}.
\begin{remark}
In order to perform pilot pattern adaptation and user scheduling
following the guidelines in \eqref{eq:distinct_spacing}--\eqref{eq:guide3} it is
necessary that the values of $f_g$ and $\tau_g$ for $g=1,\ldots,G$ are available
at the BS. Interestingly, acquiring these values can be achieved without
additional overhead. Indeed, variation over time of the maximum Doppler
frequency shift and the maximum delay spread of a channel is typically much
slower than the variations of the channel coefficients. These parameters can
thus be estimated based on previous uplink pilot transmissions.
\end{remark}

\subsection{Grouping Based Pilot Pattern Adaptation and Scheduling}
\label{sec:grouping_first}

The proposed scheme consists in first pre-assigning the RBs to
the groups $\mathcal{G}_1$,\dots,$\mathcal{G}_G$ using a mapping
$g_r:\{1\cdots N_{\mathrm{RB}}\}\to\{1\cdots G\}$. This mapping could be
the outcome of optimizing RB allocation to the pilot pattern groups based on
average per-RB channel quality indicators. Otherwise, $g_r$ could be a
{\it fixed} pre-assignment of RBs. One example of such mapping is the one
adopted in Algorithm~\ref{algo:grouping} and which satisfies
$\forall g\in\{1\cdots G\}$,
$\left|\left\{r|g_r=g\right\}\right|/N_{\mathrm{RB}}\approx|\mathcal{G}_g|/K$ to
guarantee fairness among the different groups.

Once this pre-assignment is done, the per-RB pilot pattern adaptation
consists in choosing the pilot pattern $\mathcal{P}(\mathcal{G}_{g_r})$ that has
the largest pilot inter symbol distances $\delta_{\mathcal{P}}^{\mathrm{SC}}$
and $\delta_{\mathcal{P}}^s$ satisfying the condition in \eqref{eq:guide2}.
Then, the scheduler chooses
$\mathcal{U}_r\subset\mathcal{G}_{g_r}$.
\begin{algorithm}
  \caption{Grouping Based Pilot Pattern Adaptation and User Scheduling
	(with fixed RB pre-assignment)}
 \label{algo:grouping}
\begin{algorithmic}
	\FOR{$g\in\{1\cdots G\}$}
		\FOR{$r\in\left\{
		\left\lceil\frac{N_{\mathrm{RB}}|\cup_{h=1}^{g-1}\mathcal{G}_h|}{K}
		\right\rceil+1,\ldots,
		\left\lceil\frac{N_{\mathrm{RB}}|\cup_{h=1}^{g}\mathcal{G}_h|}{K}
		\right\rceil\right\}$}
			\STATE $\mathcal{P}_r\leftarrow\mathcal{P}(\mathcal{G}_{g})$
			\STATE $\mathcal{U}_r\leftarrow\mathcal{U}\subset\mathcal{G}_{g}$
		\ENDFOR
 \ENDFOR
\end{algorithmic}
\end{algorithm}
The asymptotic results given below are valid for
{\it arbitrary} $\mathcal{U}_r$ including those obtained by
applying state-of-the-art scheduling paradigms to $\mathcal{G}_{g_r}$ and those
obtained by random selection of $\mathcal{U}_r$ from within $\mathcal{G}_{g_r}$.
\begin{remark}
Because of the grouping step and the possibility of arbitrarily choosing
$\mathcal{U}_r\subset\mathcal{G}_{g_r}$, Algorithm~\ref{algo:grouping} is much
less demanding in both computational complexity and CSI acquisition overhead
than any conventional scheme that tries to solve \eqref{eq:max_conv}.
Furthermore, we show that Algorithm~\ref{algo:grouping} outperforms
any conventional scheme that uses fixed pilot pattern assignment, at least for
large-enough numbers of users and BS antennas. As for the signaling overhead
needed to inform a user of the selected pilot pattern, it is of the order of
$\log G$ which is typically very small, e.g. only 2 bits are needed when $G=4$.
\end{remark}
We focus on the case where perfect CSI\footnote{In practice, this case amounts
to assuming that the MSE of uplink channel estimation is negligible and that
channel aging is  {\it not} an issue for downlink transmission. Similar results
can be obtained in the case of imperfect CSI and/or for other combining and
precoding criteria but are not included.} is available at the BS
and where the combining coefficients $\mathbf{w}_{k,r,t,n}^{\mathrm{UL}}$ are
chosen based on the MRC criterion and the precoding coefficients
$\mathbf{w}_{k,r,t,n}^{\mathrm{DL}}$ based on the MRT criterion. In this case,
let $R^{\mathrm{grp}}\stackrel{\mathrm{def}}{=}\frac{1}{N_{\mathrm{RB}}}
\sum_{r=1}^{N_{\mathrm{RB}}}
R_{r}\left(\mathcal{U}_r,\mathcal{P}(\mathcal{G}_{g_r})\right)$ be the spectral
efficiency achieved by Algorithm~\ref{algo:grouping}, where $R_r(.,.)$ is
given by~\eqref{eq:general_R} and where $\mathrm{grp}$ stands for `grouping'.
The following theorem states that Algorithm~\ref{algo:grouping} asymptotically
outperforms any conventional pilot pattern selection and user scheduling in
terms of average spectral efficiency for a sufficiently large number of antennas
at the BS and of users in the cell. 
\begin{theo}
 \label{theo:asym_grouping}
Assume that $\forall r$, the empirical distribution of the large-scale fading
coefficients $\{\eta_k\}_{k\in\mathcal{U}_r}$converges as
$U^{\mathrm{mux}}\to\infty$ to the distribution of a random variable $\eta$ with
mean $\overline{\eta}$. Then as $M$, $U^{\mathrm{mux}}$, $N_{\mathrm{RE}}$,
$|\mathcal{G}_g|\to\infty$ such that $U^{\mathrm{mux}}/M\to\alpha$,
$U^{\mathrm{mux}}/N_{\mathrm{RE}}\to\beta$, $|\mathcal{G}_g|/K\to\gamma_g$
where $\alpha$, $\beta$, $\gamma_g$ are constants,
\begin{equation}
 \label{eq:asym_ineq}
 \lim_{M,U^{\mathrm{mux}}\to\infty}\mathbb{P}\left\{
R^{\mathrm{grp}}>
R^{\mathrm{conv}}\right\}=1.
\end{equation}
\end{theo}
We let $N_{\mathrm{RE}}$ (which is fixed in practice) grow with
$U^{\mathrm{mux}}$ only to get nontrivial asymptotic expressions since
$|\mathcal{P}(\mathcal{G}_g)|$ also grows with $U^{\mathrm{mux}}$ due to
\eqref{eq:multiple_mux}. The assumption about the empirical distribution of
$\{\eta_k\}_{k\in\mathcal{U}_r}$ for all $r$ is also technical and is in
practice satisfied in any cell with a sufficiently large number of users that
are randomly distributed over the cell area. In this case, roughly speaking,
even the baseline scheduler employing exhaustive search ends up assigning to
each RB $r$ a set $\mathcal{U}_r$ of users which have diverse pathloss profiles,
thus validating the assumption.
\begin{proof}
For given empirical values $\{\eta_k\}_{k\in\mathcal{K}}$, the tools
of~\cite[Theorem~3]{ul_dl} can be applied to $\mathrm{SINR}_{k,r,t,n}$
defined by \eqref{eq:uplink_sinr} and \eqref{eq:downlink_sinr} to show that
$\forall r,t,n,\mathcal{U}_r$,
$\mathrm{SINR}_{k,r,t,n}-\mathrm{SINR}_{k,r}^{\det}
\stackrel{a.s.}{\rightarrow}0$, where $\det$ stands for
`deterministic equivalent' and where
\begin{equation}
 \label{eq:emp_sinr}
 \begin{multlined}
 \mathrm{SINR}_{k,r}^{\det}=\\
 \left\{
 \begin{array}{ll}
  \frac{\eta_k P^{\mathrm{UL}}}{\sigma^2/M+
(1/M)\sum_{j\in\mathcal{U}_r}\eta_j P^{\mathrm{UL}}},& \textrm{in the uplink,}\\
	\frac{\eta_k P^{\mathrm{DL}}}{\sigma^2/M+
(U^{\mathrm{mux}}/M)\eta_k P^{\mathrm{DL}}},& \textrm{in the downlink.}
 \end{array}
 \right.
 \end{multlined}
\end{equation}

Next, by our assumption about the empirical distribution of
$\eta_k$, we can apply the continuous-mapping theorem along with standard
convergence arguments to show after some tedious, but rather straightforward,
steps that \eqref{eq:general_R} and \eqref{eq:emp_sinr} lead to
\begin{align}
 \frac{R^{\mathrm{grp}}}{U^{\mathrm{mux}}}-\sum_{g=1}^{G}
\frac{|\mathcal{G}_g|}{K}
\left(1-\frac{|\mathcal{P}(\mathcal{G}_g)|}{N_{\mathrm{RE}}}\right)
\log\left(1+\overline{\mathrm{SINR}}\right)
&\stackrel{p}{\rightarrow}0,\label{eq:mrc_asym_grouping}\\
 \frac{R^{\mathrm{conv}}}{U^{\mathrm{mux}}}-
\left(1-\frac{\max_{g}|\mathcal{P}(\mathcal{G}_g)|}{N_{\mathrm{RE}}}\right)
\log\left(1+\overline{\mathrm{SINR}}\right)
&\stackrel{p}{\rightarrow}0,\label{eq:mrc_asym_conventional}
\end{align}
where
\begin{equation}
 \begin{multlined}
  \overline{\mathrm{SINR}}\stackrel{\mathrm{def}}{=}\\
  \left\{
	 \begin{array}{ll}
	  2^{\mathbb{E}_{\eta}\left[\log\left(1+
		\frac{\eta P^{\mathrm{UL}}}{\sigma^2/M+\left(U^{\mathrm{mux}}/M\right)
		\overline{\eta}P^{\mathrm{UL}}}\right)\right]}-1,& \textrm{in the uplink,}\\
		2^{\mathbb{E}_{\eta}\left[\log\left(1+
		\frac{\eta P^{\mathrm{DL}}}{\sigma^2/M+\left(U^{\mathrm{mux}}/M\right)
		\eta P^{\mathrm{DL}}}
		\right)\right]}-1,& \textrm{in the downlink.}
	 \end{array}
	\right.
 \end{multlined}
\end{equation}
Finally, since $\sum_{g=1}^{G}\frac{|\mathcal{G}_g|}{K}=1$ and
$\max_{h}|\mathcal{P}(\mathcal{G}_h)| \geq |\mathcal{P}(\mathcal{G}_g)|$
$\forall g\in\{1\cdots G\}$, from \eqref{eq:mrc_asym_grouping} and
\eqref{eq:mrc_asym_conventional} we get \eqref{eq:asym_ineq}.
\end{proof}

\section{Numerical Results}
\label{sec:simus}

We evaluate the spectral efficiency of the proposed pilot allocation scheme and
we compare it to that obtained by conventional pilot allocation.
The performance is computed assuming $N_{\mathrm{RB}}=4$ RBs, to each of which
$U^{\mathrm{mux}}$ users are allocated from a total of
$K=N_{\mathrm{RB}}U^{\mathrm{mux}}$ users.
We let $U^{\mathrm{mux}}$ vary from $4$ to $7$.
We consider $G=4$ possible profiles of time-frequency second-order
statistics characterizing the users' channels. The values for the different
Doppler frequencies and delay spreads are taken from~\cite{3gpp_etu_epa} and are
summarized in Table~\ref{table:parameters_channels}. 
\begin{table}[h]
 \caption{Channel parameters for $G=4$ groups.}
 \centering
 \begin{tabular}{| c| c| c| c|}
							\hline
						  Group index &Model &Doppler shift &Delay spread\\
							$g$ &Name &$f_g$ (Hz) &$\tau_g$ ($\mu$s)\\
							\hline
							$1$ & EPA5  & $5$ & $0.41$ \\ 
							\hline 
							$2$ & EVA70 & $70$ & $2.51$ \\
							\hline  
							$3$ & ETU70 & $70$ & $4.69$\\
							\hline 
							$4$ & ETU300 & $300$ & $4.69$\\
							\hline 
 \end{tabular}
 \label{table:parameters_channels}
\end{table}
In the table, EPA stands for the ``Extended Pedestrian A'', EVA for the
``Extended Vehicular A'' and ETU for the ``Extended Type Urban'' channel models.
Groups $\mathcal{G}_1$, $\mathcal{G}_2$, $\mathcal{G}_3$ and $\mathcal{G}_4$,
each of which composed of users that share the same channel profile, are assumed
to all have the same size: $|\mathcal{G}_g|=K/G$, $\forall g=1,2,3,4$.
Finally, $\forall k\in\mathcal{K}$,
the SNR $\eta_k P^{\mathrm{UL/DL}}/\sigma^2=10$ dB.
At each random channel realization, the SINR is evaluated per user and per RE in
each RB according to \eqref{eq:uplink_sinr} and \eqref{eq:downlink_sinr}.
When the conventional pilot pattern is used, the $K$ users are
scheduled to the $4$ RBs based on \eqref{eq:max_conv}. In order to
do so, we perform an exhaustive search among all possible user allocation
combinations and we choose the one that results in the highest spectral
efficiency. Instead, when the proposed scheme is used, we only optimize the
association of the $G$ pilot pattern groups to the $N_{\mathrm{RB}}$ RBs.

\begin{figure}[h]
 \centering
  \includegraphics[width=\linewidth]{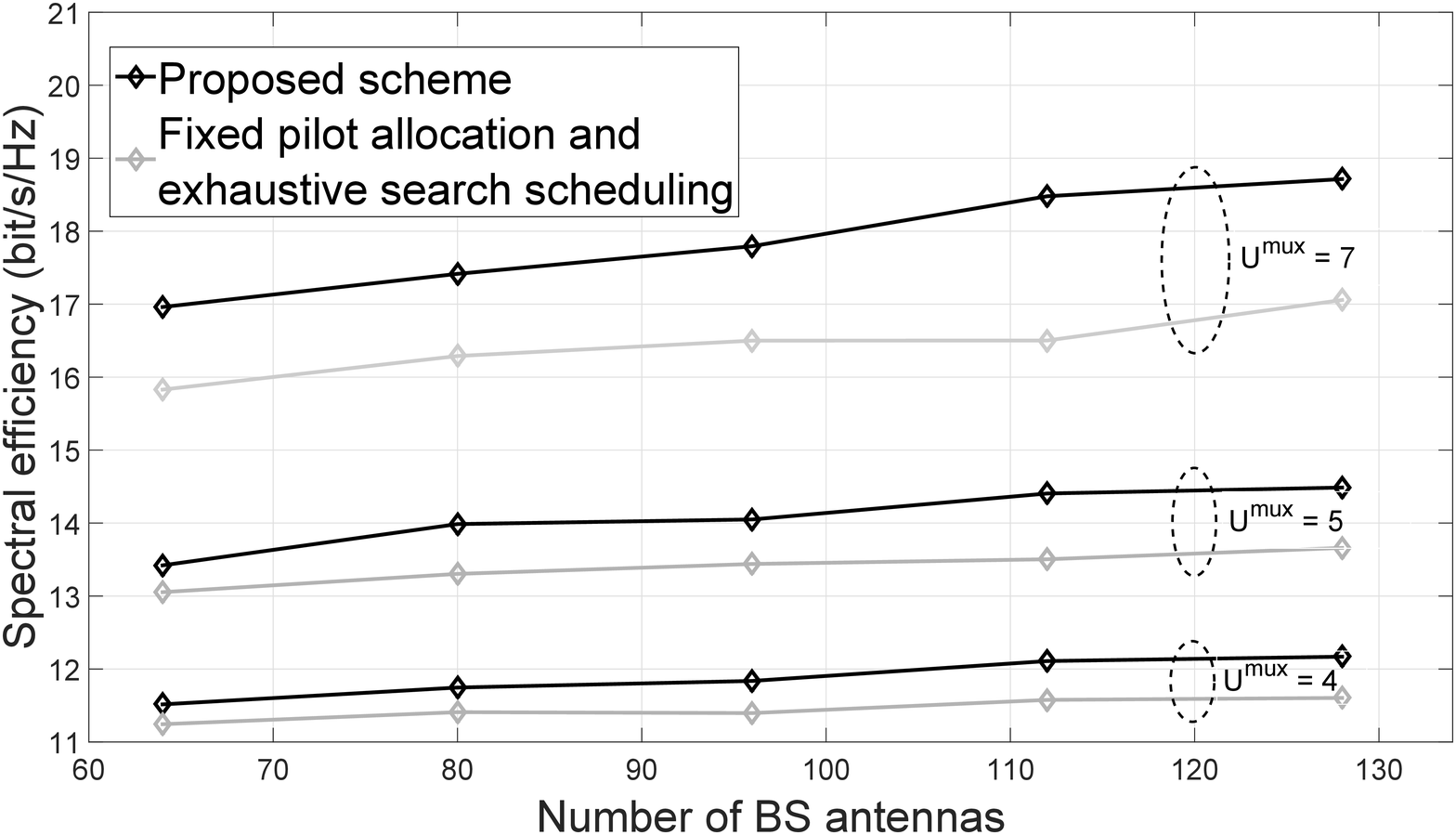}
 \caption{Average spectral efficiency vs. $M$ (SNR=10 dB).}
 \label{fig:spectralEfficiency}
\end{figure}

Fig.~\ref{fig:spectralEfficiency} shows the average spectral efficiency obtained
by the proposed scheme (black curves), and that for the conventional pilot
allocation with exhaustive search scheduling (gray curves). These results are
obtained by averaging over $10$ channel realizations. In all configurations, the
proposed scheme achieves a higher spectral efficiency, even for a moderate
number of antennas at the BS, e.g. $M=64$, and a moderate number of users per
RB, e.g. $U^{\mathrm{mux}}=4$, and in spite of the fact that exhaustive search
is performed when using the conventional scheme. This is due to the fact that,
thanks to the asymptotic channel properties of large antenna arrays, the gain in
spectral efficiency due to the increase in the number of summations in
\eqref{eq:general_R} with the proposed scheme, i.e. the increase in the average
value of $\left(1-\frac{|\mathcal{P}_r|}{N_{\mathrm{RE}}}\right)$, outweighs any
potential decrease in the term $\log\left(1+\mathrm{SINR}_{k,r,t,n}\right)$
due to restricting the scheduler with the grouping step.

\begin{figure}[h]
 \centering
  \includegraphics[width=\linewidth]{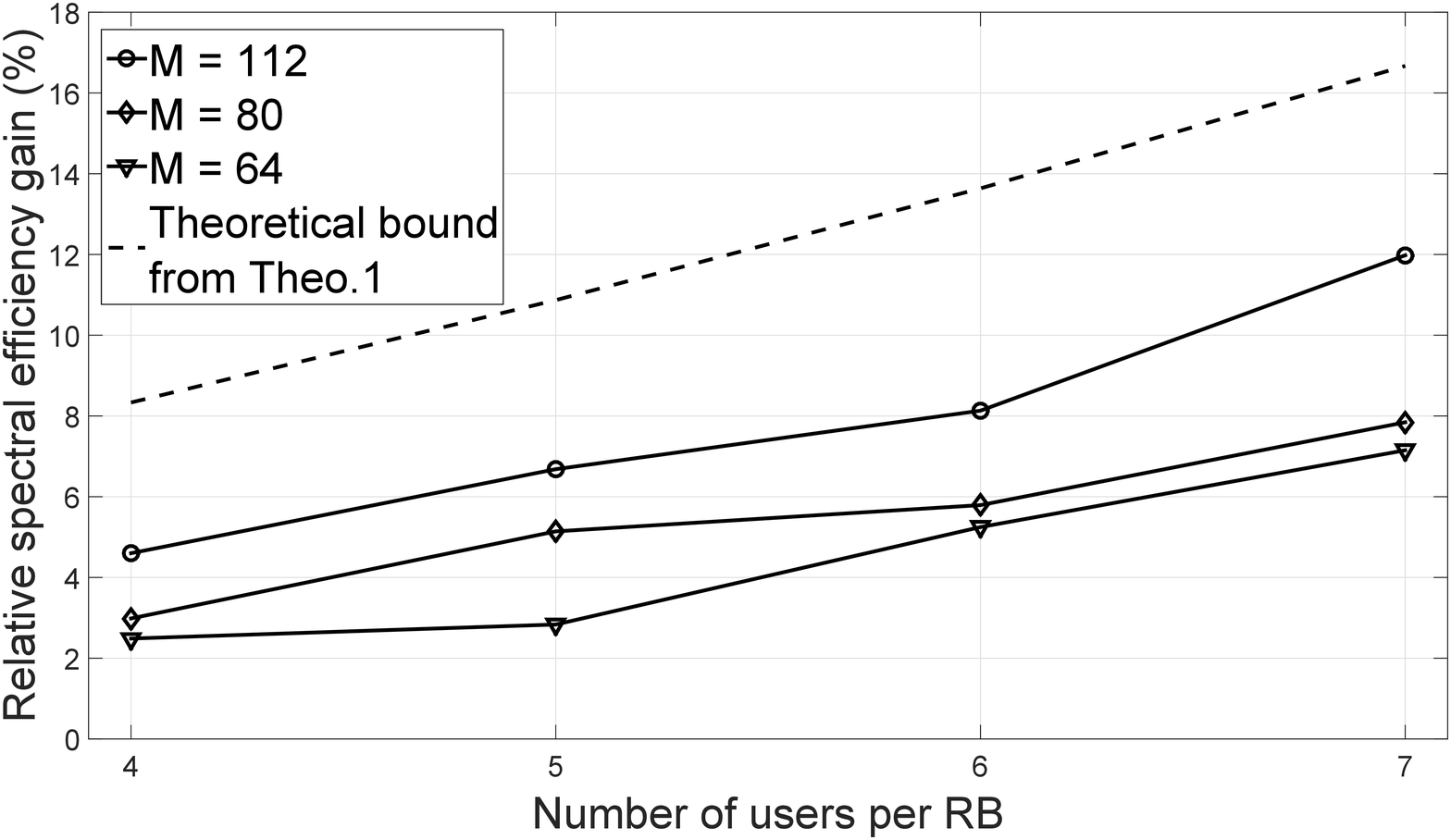}
 \caption{Spectral efficiency gain vs. $U^{\mathrm{mux}}$.}
 \label{fig:gain}
\end{figure}

Fig.~\ref{fig:gain} shows the relative gain in average spectral efficiency
w.r.t. conventional pilot assignment with exhaustive search scheduling
for different values of $M$ and $U^{\mathrm{mux}}$.
The dashed curve is the theoretical upper bound derived from
\eqref{eq:mrc_asym_grouping} and \eqref{eq:mrc_asym_conventional}. As expected,
with larger values of $M$ the gain gets closer to the asymptotic upper
bound. For instance, at $U^{\mathrm{mux}}=7$ the relative gain increases from
$7\%$ to $12\%$ when $M$ increases from 64 to 112, thus getting closer to the
$16\%$ upper bound. This gain will be even larger when practical scheduling
methods that are not based on exhaustive search are used as baseline.

\section{Conclusion}
In this paper, we presented a pilot pattern adaptation scheme for MU-MIMO that
consists in grouping users based on their pilot density requirements.
We further analytically proved that any state-of-the-art scheduling method when
used along with fixed pilot pattern assignment will be outperformed by the
proposed scheme in the limit of large numbers of users and BS antennas,
provided that users are affected by sufficiently diverse channel conditions.
We finally showed through simulations that this advantage holds even with
moderate values of these parameters. Future research directions include
proposing schemes capable of performing joint pilot pattern selection and user
scheduling and studying the effect of pilot pattern adaptation on pilot
contamination in multicell MU-MIMO scenarios.


\begin{thebibliography}{10}

\bibitem{mimo_myths}
E. Bj\"ornson, E. G. Larsson, and T. L. Marzetta, ``Massive MIMO:
Ten Myths and One Critical Question,'' \emph{IEEE Communications Magazine},
vol. 54, no. 2, pp. 114-123, February 2016.

\bibitem{beatrice_maxime}
B. Tomasi and M. Guillaud, \emph{Pilot Length
Optimization for Spatially Correlated Multi-User MIMO Channel Estimation},
in \emph{Asilomar Conference on Signals, Systems and Computers},
Pacific Grove, CA, Nov. 2015.

\bibitem{jsdm} A. Adhikary, J. Nam, J. Ahn, and G. Caire, ``Joint Spatial
Division and Multiplexing: The Large-Scale Array Regime,'' IEEE Trans.
Info. Theory, vol. 59, no. 10, pp. 6441–6463, Oct. 2013.

\bibitem{su_ofdm1}
J.-C. Guey, A. Osseiran, \emph{Adaptive Pilot Allocation in Downlink OFDM}, in
\emph{WCNC}, Las Vegas, NV, Mar. 2008.


\bibitem{su_mimo}
M. Simko, P. S. R. Diniz, and M. Rupp, \emph{Design Requirements of Adaptive
Pilot-Symbol Patterns}, in \emph{ICC}, Budapest, June 2013.

\bibitem{qos_pilot}
O. Simeone and U. Spagnolini, \emph{Adaptive Pilot Pattern for OFDM Systems},
in \emph{ICC}, Paris, June 2004.

\bibitem{speed_pilot}
S. Lee, J. Y. Lee, and H. S. Lee, \emph{Group-Based Pilot Design Method in
Mobile OFDMA Systems}, in \emph{ICACT}, Phoenix Park, Korea, Feb. 2008.

\bibitem{lte_advanced}
C. Lim, T. Yoo, B. Clerckx, B. Lee, and B. Shim, ``Recent Trend of Multiuser
MIMO in LTE-Advanced,'' IEEE Commun. Mag., vol. 51,  no. 3, pp. 127-135, Mar.
2013.

\bibitem{pilot_rule}
P. Hoeher, S. Kaiser, and P. Robertson, \emph{Two-Dimensional Pilot-Symbol-Aided
Channel Estimation by Wiener Filtering}, in \emph{ICCASP}, Munich, Apr.
1997.

\bibitem{ul_dl}
J. Hoydis, S. t. Brink, M. Debbah, ``Massive MIMO in the UL/DL of Cellular
Networks: How Many Antennas Do We Need?'' \emph{IEEE J. Sel. Areas Commun.},
vol. 31, no. 2, pp. 160-171, Feb. 2013.


\bibitem{3gpp_etu_epa}
The 3rd Generation Partnership Project (3GPP), \emph{Evolved Universal
Terrestrial Radio Access (E-UTRA); Base Station (BS) radio transmission and
reception}. Available: http://www.3gpp.org/, Sept. 2015.


\end{thebibliography}
\end{document}